\documentclass[12pt]{article}
\usepackage{graphics}
\usepackage{bm}
\usepackage{graphicx}
\usepackage{amsmath}
\usepackage{amssymb}
\usepackage{amstext}
\usepackage{amscd}
\usepackage{amsfonts}
\usepackage{appendix}
\usepackage{wasysym}
\usepackage{textcomp}
\usepackage{epstopdf}
\usepackage{color}
\DeclareMathAlphabet{\EuFrak}{U}{euf}{m}{n}
\DeclareMathAlphabet{\EuScript}{U}{eus}{m}{n}

\newcommand{\nd}{\noindent}

\newcommand{\be}{\begin{equation}}
\newcommand{\ee}{\end{equation}}
\newcommand{\ben}{\begin{eqnarray}}
\newcommand{\een}{\end{eqnarray}}

\title{{\bf Relativistic treatment of Verlinde's emergent force
in Tsallis' statistics}}
\author{{L. Calderon$^{1,4}$, M. T. Martin$^{3,4}$, A. Plastino$^{2,4,5,6}$,}\\ 
{M. C. Rocca$^{2,3,4,5}$,V. Vampa}$^{1}$ \\
\small{$^1$Departamento de Ciencias B\'asicas, Facultad de Ingenier\'{\i}a},\\
\small{$^2$ Departamento de F\'{\i}sica,
Universidad Nacional de La Plata},\\
\small{$^3$ Departamento de Matem\'{a}tica,
Universidad Nacional de La Plata,}\\
\small{$^4$ Consejo Nacional de Investigaciones Cient\'{\i}ficas
y Tecnol\'{o}gicas, Argentina},\\
\small{$^5$IFLP-CCT-CONICET-C. C. 727, 1900 La Plata, Argentina},\\
\small{$^6$  SThAR - EPFL, Lausanne, Switzerland}}

\date{\today}

\begin{document}

\maketitle

\begin{abstract}
\nd Following Chakrabarti,Chandrasekhar, and Naina  
[Physica A {\bf 389} (2010) 1571], we attempt a classical
relativistic treatment of Verlinde's emergent entropic force conjecture by
appealing to a relativistic Hamiltonian in the context of
Tsalli's statistics. The ensuing partition function becomes the 
classical one for small velocities.
We show that Tsallis' relativistic
(classical) free particle distribution at temperature $T$ can generate
Newton's gravitational force's $r^{-2}$  {\it distance's dependence}.
If we want to repeat the concomitant argument 
by appealing to Renyi's distribution, the attempt fails and one needs to
modify the conjecture.\\
\nd Keywords: Tsallis' and Renyi's relativistic distributions,
classical partition function, entropic force.\\
\nd PACS: {05.20.-y, 05.70.Ce, 05.90.+m}

\end{abstract}

\newpage

\tableofcontents

\newpage

\renewcommand{\theequation}{\arabic{section}.\arabic{equation}}

\section{Introduction}

\setcounter{equation}{0}

In 2011, Verlinde \cite{verlinde}    put forward a conjecture that  connects gravity to an entropic force. Gravity would then arise out of information regarding the positions of material bodies (it from bit). This idea 
links a thermal gravity-treatment to 't Hooft's holographic principle. As a consequence,  gravitation ought to be be regarded as  an emergent phenomenon. Verlinde's conjecture attained considerable reception (just as an example, see \cite{times}). For a superb overview on the statistical mechanics of gravitation, we recommend Padmanabhan's work \cite{india}, and references therein.\vskip 2mm

\nd Verlinde's initiative originated works on cosmology, the dark energy hypothesis, cosmological acceleration, cosmological inflation, and loop quantum\\ gravity. The literature is immense \cite{libro}. A relevant  contribution to information theory is that of Guseo \cite{guseo}, who  proved that the local entropy function, related to a
logistic distribution, is a catenary and vice versa. Such  invariance may
be explained, at a deeper level, through the Verlinde’s conjecture on the origin
of gravity, as an effect of the entropic force. Guseo puts forward a new  interpretation of the local entropy in a system, as quantifying a hypothetical attraction force that
the system would exert  \cite{guseo}.
    \vskip 2mm

\nd The present effort does not deal with any of these issues. What we will do is to show that a  simple classical reasoning centered on Tsallis' relativistic
probability distributions  proves Varlinde's conjecture. For Renyi's relativistic
instance, one needs to modify the conjecture to achieve a similar result.

\nd Our point of departure is Ref. \cite{naina}, in which their authors studied a canonical ensemble of $N$ particles for a classical relativistic ideal gas, and found its specific heat in
the Tsallis-Mendes-Plastino (TMP) scenario \cite{tsallis}.  We will not use here the TMP scenario.  Inspired by  \cite{naina},  we  appeal as well to our previous effort \cite{epl} for non-relativistic results and deal  with Tsallis' statistics with linear constraints as a priori information \cite{tsallis}. In addition to finding, for the first time ever, relativistic Verlinde-results in a Tsallis'context,  we will, for the sake of completeness, register some advances regarding the relativistic Tsallis scenario with linear constraints for the ideal gas.

\setcounter{equation}{0}

\section{Tsallis' relativistic partition function for the free particle} 

\nd The celebrated and well-known Tsalis entropy is a generalization of Shanon's one, that depends on a free real parameter $q$ \cite{tsallis}. \vskip 3mm 
\nd {\bf The $q<1$ instance} \vskip 3mm 
\nd We consider first the case $q<1$. This case is not relevant to our Verlinde's endeavor \cite{epl}, but is a logical addition to the results of \cite{naina}. \vskip 3mm

\nd Tsallis' relativistic q-partition function for  $N-$free particles of mass $m$ 
reads \cite{naina}
\begin{equation}
\label{eq5.1}
{\cal Z}=\frac {V} {N!h^{3N}}\int \left[1+(1-q)\beta(\sqrt{m^2c^4+p^2c^2}-mc^2)
\right]_+^{\frac {1} {q-1}} d^4p.
\end{equation}  
Using spherical coordinates and integrating over the angles the precedent integral
we have
\begin{equation}
\label{eq5.2}
{\cal Z}=\frac {4\pi V} {N!h^{3N}}\int\limits_0^\infty 
\left[1+(1-q)\beta(\sqrt{m^2c^4+p^2c^2}-mc^2)
\right]^{\frac {1} {q-1}} p^2 dp.
\end{equation}  
With the change of variables  $y^2=p^2+m^2c^2$ one now has
\begin{equation}
\label{eq5.3}
{\cal Z}=\frac {4\pi V} {N!h^{3N}}\int\limits_{mc}^\infty 
y\sqrt{y^2-m^2c^2}
\left[1+(1-q)\beta c(y-mc)\right]^{\frac {1} {q-1}} dy.
\end{equation}  
Let $x$ be given by $y=mcx$. We have then
\begin{equation}
\label{eq5.4}
{\cal Z}=\frac {4\pi Vm^3c^3} {N!h^{3N}}\int\limits_1^\infty 
x\sqrt{x^2-1}
\left[1+(1-q)\beta mc^2(x-1)\right]^{\frac {1} {q-1}} dx.
\end{equation}  
With $s$ defined as $x=s+1$ we obtain:
\begin{equation}
\label{eq5.5}
{\cal Z}=\frac {4\pi Vm^3c^3} {N!h^{3N}}\int\limits_0^\infty 
\left(s^{\frac {3} {2}}+s^{\frac {1} {2}}\right)
\left(s+2\right)^{\frac {1} {2}}
\left[1+(1-q)\beta mc^2s\right]^{\frac {1} {q-1}} ds,
\end{equation}  
or 
\[{\cal Z}=\frac {4\pi Vm^3c^3} {N!h^{3N}}
[(1-q)\beta m c^2]^{\frac {1} {q-1}}
\int\limits_0^\infty 
s^{\frac {3} {2}}
\left(s+2\right)^{\frac {1} {2}}
\left[s+\frac {1} {(1-q)\beta mc^2}\right]^{\frac {1} {q-1}} ds+\]
\begin{equation}
\label{eq5.6}
\frac {4\pi Vm^3c^3} {N!h^{3N}}
[(1-q)\beta m c^2]^{\frac {1} {q-1}}
\int\limits_0^\infty 
s^{\frac {1} {2}}
\left(s+2\right)^{\frac {1} {2}}
\left[s+\frac {1} {(1-q)\beta mc^2}\right]^{\frac {1} {q-1}} ds.
\end{equation}  
Appealing to reference \cite{gra} we have now a result in terms of Hyper-geometric functions $F$ and Beta functions $B$, namely, 

\[{\cal Z}=\frac {4\pi Vm^3c^3} {N!h^{3N}}
[(1-q)\beta m c^2]^{-\frac {3} {2}}\left[
\frac {B\left(\frac {5} {2},\frac {1} {1-q}-3\right)}
{\beta mc^2(1-q)}\right.\times\]
\[F\left(-\frac {1} {2},\frac {5} {2},\frac {1} {1-q}-\frac {1} {2};
1-\frac {1} {2\beta mc^2(1-q)}\right)+\]
\begin{equation}
\label{eq5.7}
\left.B\left(\frac {3} {2},\frac {1} {1-q}-2\right)
F\left(-\frac {1} {2},\frac {3} {2},\frac {1} {1-q}-\frac {1} {2};
1-\frac {1} {2\beta mc^2(1-q)}\right)\right].
\end{equation} 
For $\beta mc^2>>1$, $mc^2>>k_BT$,    we are in the non-relativistic case and have

\begin{equation}
\label{eq5.8}
{\cal Z}=\frac {2\pi V} {N!h^{3N}}
\left[\frac {2m} {\beta(1-q)}\right]^{\frac {3} {2}}
\frac {\Gamma\left(\frac {3} {2}\right)
\Gamma\left(\frac {1} {1-q}-\frac {3} {2}\right)}
{\Gamma\left(\frac {1} {1-q}\right)}.
\end{equation}

\vskip 3mm 

\nd {\bf The case $q>1$} \vskip 3mm
\nd Let is now consider gravitationally relevant \cite{epl}  case $q>1$  . We have for the partition function
\begin{equation}
\label{eq5.9}
{\cal Z}=\frac {4\pi Vm^3c^3} {N!h^{3N}}\int\limits_0^\infty 
\left(s^{\frac {3} {2}}+s^{\frac {1} {2}}\right)
\left(s+2\right)^{\frac {1} {2}}
\left[1-(q-1)\beta mc^2s\right]_+^{\frac {1} {q-1}} ds.
\end{equation}  
Integrating on the angles we have again
\begin{equation}
\label{eq5.10}
{\cal Z}=\frac {4\pi Vm^3c^3} {N!h^{3N}}
\int\limits_0^{\frac {1} {\beta mc^2(q-1)}}
\left(s^{\frac {3} {2}}+s^{\frac {1} {2}}\right)
\left(s+2\right)^{\frac {1} {2}}
\left[1-(q-1)\beta mc^2s\right]^{\frac {1} {q-1}} ds,
\end{equation}  
or 
\[{\cal Z}=\frac {4\pi Vm^3c^3} {N!h^{3N}}
[(q-1)\beta m c^2]^{\frac {1} {q-1}}
\int\limits_0^{\frac {1} {\beta mc^2(q-1)}}
s^{\frac {3} {2}}
\left(s+2\right)^{\frac {1} {2}}
\left[\frac {1} {(q-1)\beta mc^2}-s\right]^{\frac {1} {q-1}} ds+\]
\begin{equation}
\label{eq5.11}
\frac {4\pi Vm^3c^3} {N!h^{3N}}
[(q-1)\beta m c^2]^{\frac {1} {q-1}}
\int\limits_0^{\frac {1} {\beta mc^2(q-1)}}
s^{\frac {1} {2}}
\left(s+2\right)^{\frac {1} {2}}
\left[\frac {1} {(q-1)\beta mc^2}-s\right]^{\frac {1} {q-1}} ds.
\end{equation}  
By recourse to \cite{gra} we now obtain
\[{\cal Z}=\frac {2\pi V} {N!h^{3N}}
\left[\frac {2m} {\beta m(q-1)}\right]^{\frac {3} {2}}
\left[\frac {B\left(\frac {5} {2},\frac {1} {q-1}+1\right)}
{\beta mc^2(q-1)}\right.\times\]
\[F\left(-\frac {1} {2},\frac {5} {2},\frac {7} {2}+\frac {1} {q-1};
-\frac {1} {2\beta mc^2(q-1)}\right)+\]
\begin{equation}
\label{eq5.12}
\left.B\left(\frac {3} {2},\frac {1} {q-1}+1\right)
F\left(-\frac {1} {2},\frac {3} {2},\frac {5} {2}+\frac {1} {q-1};
-\frac {1} {2\beta mc^2(q-1)}\right)\right].
\end{equation} 
For $\beta mc^2>>1$, the classic case, the partition function reads
\begin{equation}
\label{eq5.13}
{\cal Z}=\frac {2\pi V} {N!h^{3N}}
\left[\frac {2m} {\beta(q-1)}\right]^{\frac {3} {2}}
\frac {\Gamma\left(\frac {3} {2}\right)
\Gamma\left(\frac {1} {q-1}+1\right)}
{\Gamma\left(\frac {1} {1-q}+\frac {5} {2}\right)},
\end{equation}
which is the usual non relativistic Tsalli's partition function for $q>1$ already obtained in 
 \cite{epl}.
Figure 1 displays the graph of the function $ H(T)$ given by
\begin{equation}
\label{eq5.14}
{\cal Z}=\frac {2\pi V} {N!h^{3N}}
\left[\frac {2m} {\beta(q-1)}\right]^{\frac {3} {2}}
H(T),
\end{equation}
for $q=\frac {4} {3}$, the specific $q-$value needed for gravitaional considerations \cite{epl}.
It tells us that ${\cal Z}$ is always positive, as it should be.

\newpage

\begin{figure}[h]
\begin{center}
\includegraphics[scale=0.5,angle=0]{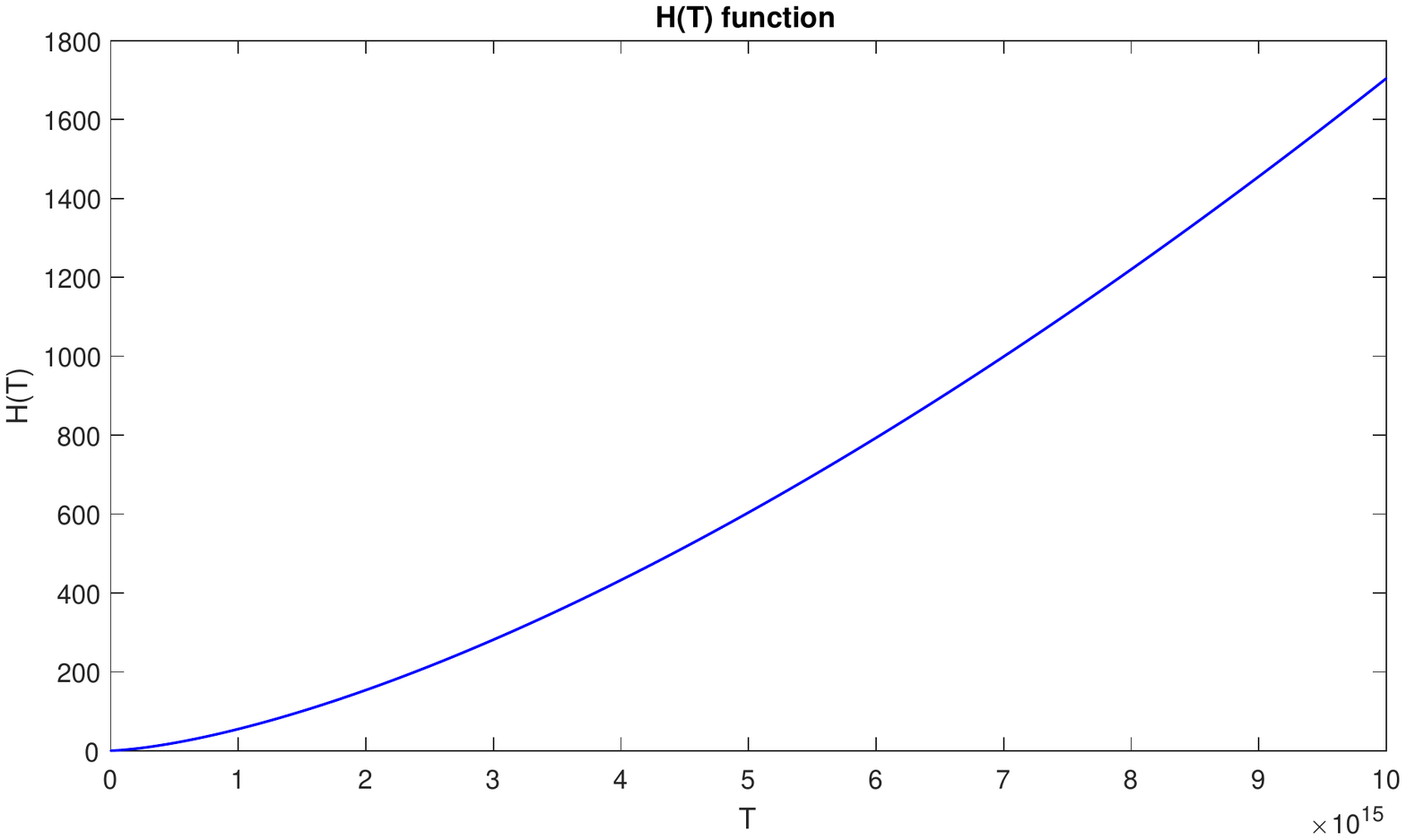}
\vspace{-0.2cm} \caption{H function}\label{fig1}
\end{center}
\end{figure}

\newpage

\section{Tsallis'  relativistic mean energy of the free particle}
\vskip 3mm
\nd {\bf Case $q<1$} \vskip 3mm
\nd Let  us now calculate the average energy corresponding, firstly in the case $q<1$. 
For it we have
\[{\cal Z}<{\cal U}>=\frac {V} {N!h^{3N}}\int 
[\sqrt{m^2c^4+p^2c^2}-mc^2]\times\]
\begin{equation}
\label{eq5.15}
\left[1+(1-q)\beta(\sqrt{m^2c^4+p^2c^2}-mc^2)
\right]_+^{\frac {1} {q-1}} d^4p,
\end{equation}  
or 
\[{\cal Z}<{\cal U}>=\frac {V} {N!h^{3N}}\int 
[\sqrt{m^2c^4+p^2c^2}]\times\]
\begin{equation}
\label{eq5.16}
\left[1+(1-q)\beta(\sqrt{m^2c^4+p^2c^2}-mc^2)
\right]_+^{\frac {1} {q-1}} d^4p-mc^2{\cal Z}.
\end{equation}  
With changes in the variables similar to those made for the partition function, we obtain here
\[{\cal Z}<{\cal U}>=\frac {4\pi Vm^4C^5} {N!h^{3N}}\int\limits_0^{\infty}
x^{\frac {3} {2}}(x+1)(\sqrt{x+2}\times\]
\begin{equation}
\label{eq5.17}
\left[1+(1-q)\beta mc^2x
\right]^{\frac {1} {q-1}} dx.
\end{equation}  
This last equation can be rewritten as
\[{\cal Z}<{\cal U}>=\frac {4\pi Vm^4C^5} {N!h^{3N}}
[\beta m c^2(1-q)]^{\frac {1} {q-1}}
\int\limits_0^{\infty}
x^{\frac {3} {2}}(x+1)(\sqrt{x+2}\times\]
\begin{equation}
\label{eq5.18}
\left[x+\frac {1} {(1-q)\beta mc^2}
\right]^{\frac {1} {q-1}} dx.
\end{equation}  
Returning again to  reference \cite{gra}, we obtain for $<{\cal U}>$
\[<{\cal U}>=\frac {\sqrt{2}\;4\pi Vm^4c^5} {N!h^{3N}{\cal Z}}
\left[\frac {1} {\beta mc^2(1-q)}\right]^{\frac {5} {2}+\frac {1} {q-1}}
\left[\frac {B\left(\frac {7} {2},\frac {1} {1-q}-4\right)}
{\beta mc^2(1-q)}\right.\times\]
\[F\left(-\frac {1} {2},\frac {7} {2},\frac {1} {1-q}-\frac {1} {2};
1-\frac {1} {2\beta mc^2(1-q)}\right)+\]
\begin{equation}
\label{eq5.19}
\left.B\left(\frac {5} {2},\frac {1} {1-q}-3\right)
F\left(-\frac {1} {2},\frac {5} {2},\frac {1} {1-q}-\frac {1} {2};
1-\frac {1} {2\beta mc^2(1-q)}\right)\right].
\end{equation} 
From this last equation we obtain the mean energy expression for the non-relativistic case
\begin{equation}
\label{eq5.20}
<{\cal U}>=\frac {3} {\beta[2-5(1-q)]}.
\end{equation}

\vskip 3mm 

\nd {\bf Case $q$ larger than one} \vskip 3mm
When $q>1$ we have
\[{\cal Z}<{\cal U}>=\frac {4\pi Vm^4C^5} {N!h^{3N}}\int\limits_0^{\infty}
x^{\frac {3} {2}}(x+1)(\sqrt{x+2}\times\]
\begin{equation}
\label{eq5.21}
\left[1-(q-1)\beta mc^2x
\right]_+^{\frac {1} {q-1}} dx-mc^2{\cal Z}.
\end{equation}  
Making a similar reasoning as for the case $q<1$ we obtain
\[<{\cal U}>=\frac {\sqrt{2}\;4\pi Vm^4c^5} {N!h^{3N}{\cal Z}}
\left[\frac {1} {\beta mc^2(q-1)}\right]^{\frac {1} {q-1}}
\left[\frac {B\left(\frac {7} {2},\frac {1} {q-1}+1\right)}
{\beta mc^2(q-1)}\right.\times\]
\[F\left(-\frac {1} {2},\frac {7} {2},\frac {1} {q-1}+\frac {9} {2};
\frac {1} {2\beta mc^2(q-1)}\right)+\]
\begin{equation}
\label{eq5.22}
\left.B\left(\frac {5} {2},\frac {1} {q-1}+1\right)
F\left(-\frac {1} {2},\frac {5} {2},\frac {1} {q-1}+\frac {7} {2}
-\frac {1} {2\beta mc^2(q-1)}\right)\right].
\end{equation} 
For  $\beta m c^2>>1$ (the non-relativistic case) we obtain the result of  \cite{epl}, i.e., 
\begin{equation}
\label{eq5.23}
<{\cal U}>=\frac {3} {\beta[2+5(q-1)]}.
\end{equation}

\section{Specific heat in the linear constraints Tsallis' scenario}

 Let is now calculate the specific heat for the case $q=\frac {4} {3}$, relevant for Verlinde-endeavors \cite{epl}. This was not done in \cite{naina}.
We should first note, with respect to Hyper-geometric functions, that
\begin{equation}
\label{eq5.24}
\frac {d} {dz}F(\alpha,\beta,\gamma;z)=
-\alpha\beta F(\alpha+1,\beta+1,\gamma+1;z).
\end{equation}
We now use the notation
\begin{equation}
\label{eq5.25}
F_1=F\left(-\frac {1} {2},\frac {7} {2},\frac {9} {2}+3;-\frac {3k_BT} {2mc^2}\right),
\end{equation}
\begin{equation}
\label{eq5.26}
F_2=F\left(-\frac {1} {2},\frac {5} {2},\frac {7} {2}+3;-\frac {3k_BT} {2mc^2}\right),
\end{equation}
\begin{equation}
\label{eq5.27}
F_3=F\left(-\frac {1} {2},\frac {3} {2},\frac {5} {2}+3;-\frac {3k_BT} {2mc^2}\right),
\end{equation}
\begin{equation}
\label{eq5.28}
F_4=F\left(\frac {1} {2},\frac {9} {2},\frac {9} {2}+4;-\frac {3k_BT} {2mc^2}\right),
\end{equation}
\begin{equation}
\label{eq5.29}
F_5=F\left(\frac {1} {2},\frac {7} {2},\frac {7} {2}+4;-\frac {3k_BT} {2mc^2}\right),
\end{equation}
\begin{equation}
\label{eq5.30}
F_6=F\left(\frac {1} {2},\frac {5} {2},\frac {5} {2}+4;-\frac {3k_BT} {2mc^2}\right).
\end{equation}
Thus, we can write
\begin{equation}
\label{eq5.31}
<{\cal U}>=3k_BT\frac {\frac {3k_BT} {mc^2}B\left(\frac {7} {2},4\right)F_1+
B\left(\frac {5} {2},4\right)F_2} 
{\frac {3k_BT} {mc^2}B\left(\frac {5} {2},4\right)F_2+
B\left(\frac {3} {2},4\right)F_3},
\end{equation}
and, for the specific heat we have then
\[C=\frac {\partial<{\cal U}>} {\partial T}=
\frac {<{\cal U}>} {T}+\frac {9k_B^2T} {mc^2}
\frac{B\left(\frac {7} {2},4\right)F_1-
\frac {21k_BT} {3mc^2}B\left(\frac {7} {2},4\right)F_4-
\frac {5} {8}
B\left(\frac {5} {2},4\right)F_5} 
{\frac {3k_BT} {mc^2}B\left(\frac {5} {2},4\right)F_2+
B\left(\frac {3} {2},4\right)F_3}- \]
\begin{equation}
\label{eq5.32}
\frac {3k_B<{\cal U}>} {mc^2}
\frac{B\left(\frac {5} {2},4\right)F_2-
\frac {15k_BT} {8mc^2}B\left(\frac {5} {2},4\right)F_5-
\frac {3} {8}
B\left(\frac {3} {2},4\right)F_6} 
{\frac {3k_BT} {mc^2}B\left(\frac {5} {2},4\right)F_2+
B\left(\frac {3} {2},4\right)F_3}. 
\end{equation}
This expression is plotted in Figure 2. We see that
the specific heat is always positive, as it happens in the
non-relativistic case   \cite{epl}.

\newpage

\begin{figure}[h]
\begin{center}
\includegraphics[scale=0.5,angle=0]{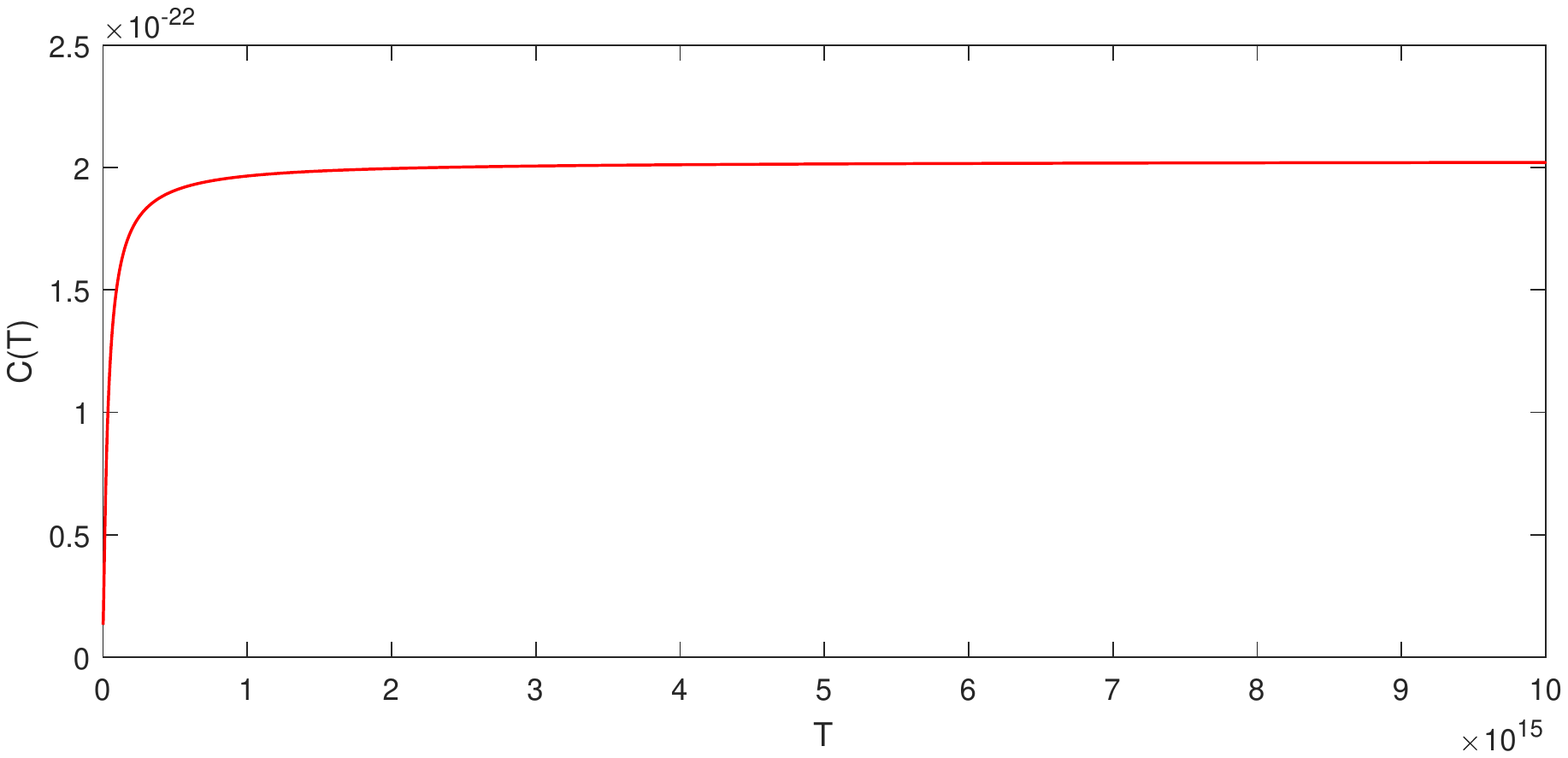}
\vspace{-0.2cm} \caption{Specific heat}\label{fig2}
\end{center}
\end{figure}

\newpage

\section{The relativistic, Tsallis entropic force}

\setcounter{equation}{0}

We arrive now at our main present goal.
We specialize things now to
$q=\frac {4} {3}$.
Why do we select this special value $q=\frac {4} {3}$? There is a solid reason. This is because
$${\cal S}=\ln_q{\cal Z}+{\cal Z}^{1-q}\beta<{\cal U}>.$$
Since the entropic force is to  be defined as proportional to the gradient of
${\cal S}$, there is a unique $q$-value  for which
the dependence on $r$ of the entropic force is $\sim r^{-2}$
when $\nu=3$. Thus we obtain, for $q=4/3$,
\begin{equation}
\label{eq6.1}
{\cal S}=3-(3-\beta<{\cal U}>){\cal Z}^{-\frac {1} {3}}.
\end{equation}
From (\ref{eq5.12}) we can write
\begin{equation}
\label{eq6.2}
<{\cal Z}>=ar^3,
\end{equation}
from which it is obtained that
\begin{equation}
\label{eq6.3}
{\cal S}=3-\frac {3-\beta<{\cal U}>}{a^{\frac {1} {3}}r}.
\end{equation}
Following Verlinde \cite{verlinde} we define the entropic force as
\begin{equation}
\label{eq6.4}
{\vec {\cal F}}_e=-\frac {\lambda} {\beta}{\vec {\nabla}{\cal S}},
\end{equation}
where $\vec{\nabla}$ indicates the four-gradient in Minkowskian space.
\begin{equation}
\label{eq6.5}
{\vec {\cal F}}_e=-\frac {\lambda} {\beta}
\frac {3-\beta<{\cal U}>}{a^{\frac {1} {3}}r^2}{\vec e}_r,
\end{equation}
where    
 ${\vec e}_r$ is the radial unit vector.
We see that $F_e$ acquires an appearance quite similar to that of Newton's gravitational one, 
as conjectured by Verlinde en \cite{verlinde}. 
In Figures 3 and 4 the function $ L=3-\beta<{\cal U}>$ is plotted. We see that $L$ is always positive. 
This entails  that the relativistic entropic force is purely gravitational.

\newpage

\begin{figure}[h]
\begin{center}
\includegraphics[scale=0.5,angle=0]{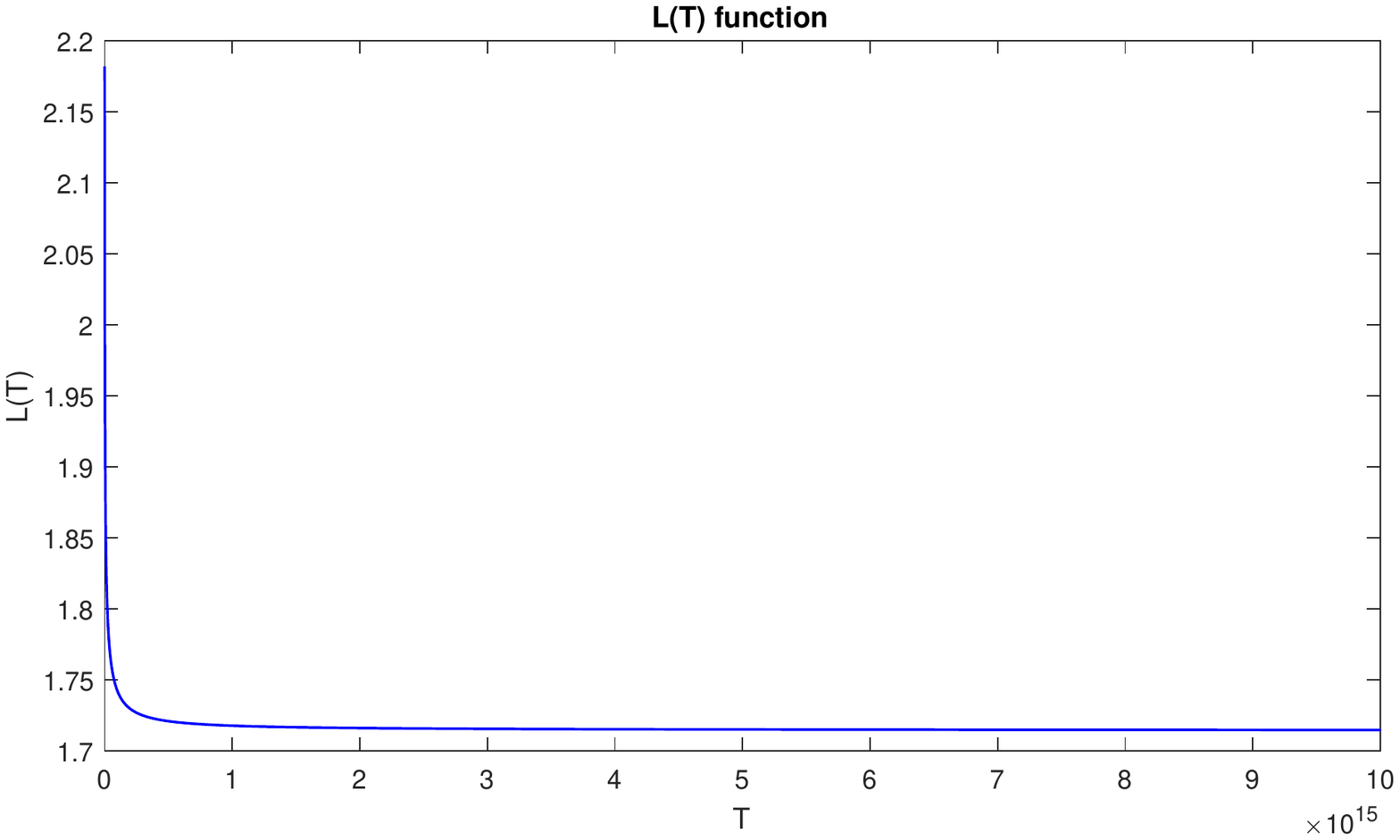}
\vspace{-0.2cm} \caption{$L(T)=3-\beta<{\cal U}>$}\label{fig3}
\end{center}
\end{figure}

\begin{figure}[h]
\begin{center}
\includegraphics[scale=0.5,angle=0]{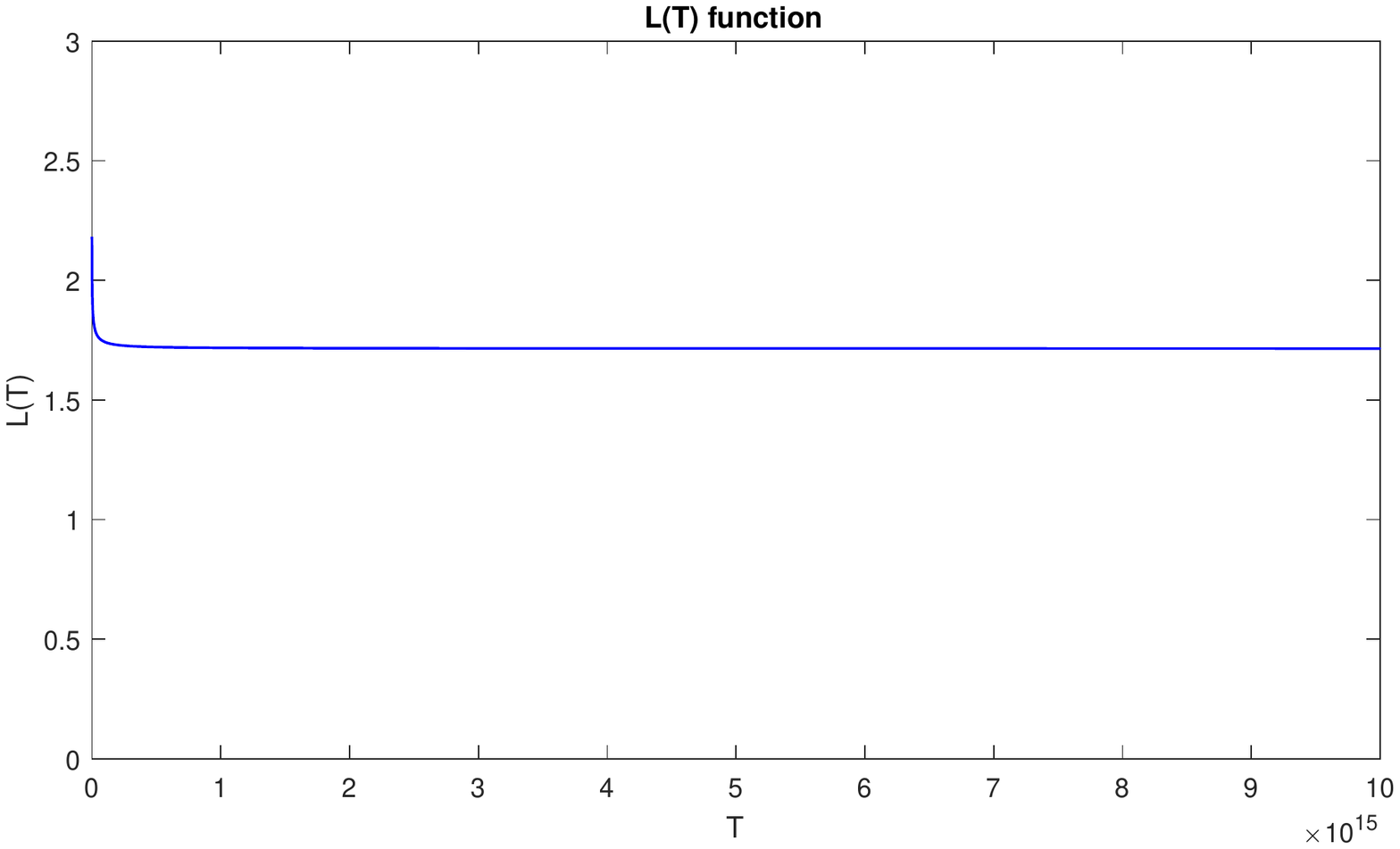}
\vspace{-0.2cm} \caption{Centered $L(T)=3-\beta<{\cal U}>$}\label{fig4}
\end{center}
\end{figure}

\section{The relativistic, Renyi's entropic force}

\setcounter{equation}{0}
In  Renyi's approach to our problem \cite{epl} the entropy is
\begin{equation}
\label{eq8.1}
{\cal S}=\ln{\cal Z}+\ln[1+(1-\alpha)\beta<{\cal U}>]_+^{\frac {1} {1-\alpha}}.
\end{equation}
For $\alpha=\frac {4} {3}$, the expression for the entropy is
\begin{equation}
\label{eq8.2}
{\cal S}=\ln{\cal Z}+\ln\left[1-\frac {\beta<{\cal U}>} {3}\right]_+^{-3}.
\end{equation}
The second term on the right hand of
(\ref{eq8.2}) is independent  of $r$. Additionally, from (\ref{eq6.2}) we obtain
\begin{equation}
\label{eq8.3}
\ln{\cal Z}=3\ln r+\ln a.
\end{equation}
  Here we need to derive the entropy with respect to the area, thus changing Verlinde´s conjecture.  As in the non-relativistic case \cite{epl}, we have then
\begin{equation}
\label{eq8.4}
{\vec F}_e=-\frac {\lambda} {\beta}
\frac {\partial{\cal S}} {\partial A}{\vec e}_R=-
\frac {\lambda} {\beta}
\frac {3} {8\pi r^2}{\vec e}_r.
\end{equation}
This is again a gravitational expression for the entropic force.

\section{Conclusions}

\nd We obtained here  the relativistic partition function ${\cal Z}$ of Tsalli's theory with linear constraints,
that adequately reduces itself  to its non-relativistic counterpart for small velocities.

\nd We do the same for the mean value of the energy$<{\cal U}>$ for
the relativistic Hamiltonian of the ideal gas.

\nd We obtain the associated specific heat that turns out to be positive,  
as befits an  ideal gas.

\nd From ${\cal Z}$ and $<{\cal U}>$ we obtained the relativistic entropy
${\cal S}$

\nd We have presented two very simple relativistic classical realizations of Verlinde's conjecture. 
The Tsallis treatment, for $q=4/3$, seems to be neater, as the entropic force is directly associated 
to the gradient of
Tsallis' entropy $S_q$, which acts as a ''potential'', as Verlinde prescribes. This is not so in the 
Renyi instance, in which one has to modify Verlinde's $F_e$ definition and derive $S$ with respect to the area.

\nd Strictly speaking, Verlinde's conjecture can be unambiguously
proved for the Tsallis entropy with $q=4/3$. The
Renyi demonstration correspond to a modified version of Verlinde's
conjecture.

\nd Of course, ours is a very preliminary, if significant,  effort.  
A much more elaborate treatment would be desirable.

\end{document}